# Design and Experimental Characterization of Spiral Resonator variants-TTSR , NBSR.

Sougata Chatterjee [#1] , Shantanu Das[*2]

[#1] NCRA-TIFR (Tata Institute for Fundamental Research) -Pune, India.
[*2] Reactor Control Division, BARC (Bhabha Atomic Research Center) Mumbai, India

*Abstract*- **In this paper we have carried out numerical simulation and experimental characterization of two variants of SR (Spiral resonator) viz. TTSR (Two-Turn Spiral Resonator) and NBSR (Non Bianisotropic spiral resonator). The numerical simulation of these develop magnetic inclusion structures are done with commercially available Finite Element Method (FEM) based Ansys HFSS-13 simulation software, and transmission experiment using S-parameter retrieval technique. The experimental characterization of those fabricated structures is used in the parallel plate waveguide experiment (not reported earlier) and then validate our experimental results with the Ansys HFSS-13 simulation software; which in not available in any published literature.**

## I. INTRODUCTION

The left handed material (LHM) exhibits counterintuitive phenomena like a reversal of Snell's law, revered Doppler effect and reversed Cerenkov radiation. The possibilities of such material were first proposed by V. G. Veselago [1], where he considered permittivity and permeability to be simultaneously negative. Materials exhibiting such counterintuitive phenomena are not available in nature and are thus known as meta-material, meaning not available in nature. Some typical characteristics are also discussed by Shantanu et al [2-6] , where the author discusses photon behavior inside a metamaterial. The possibility of realization of negative permittivity at microwave frequency was reported [7] in 1996 using thin metallic wire array, while the realization of negative permeability was subsequently reported by the same researcher J. B. Pendry et al. [8] in 1999. They proposed two structures viz: split ring resonator (SRR) and swiss-roll; the planer version of the swiss roll is called spiral resonator (SR), while J. D. Baena et al. [9] proposed two variants of SR structure: the two turn spiral resonator (TTSR) and, non bi-anisotropic spiral resonator (NBSR).

In this paper, we have done the numerical simulations of SR-variants TTSR, NBSR using commercially available FEM based simulation software (Ansys HFSS-13). All the two structures (TTSR, NBSR) have been fabricated on the basis of numerical simulated data and then fabricated structures have been characterized experimentally. The performance characteristics of TTSR and NBSR have been compared on the basis of Ansys HFSS-13 simulation and experimental result, which shows a good match between them. Though the numerical simulation is not new but experimentally characterized is the novelty of this paper. Many researchers give how to characterize metamaterial but they do not give any proper description on how this characterization process is working. The first part of this paper will discuss the simulation and experimental studies on SR variants and the next part of this paper will discuss the interpretation of the results, followed by conclusions and references.

## II. SIMULATION AND EXPERIMENTAL STUDIES OF TTSR, AND NBSR.

The unit cell structures of TTSR and NBSR are shown in Fig.1 where the magnetic field is normal to the plane of the unit cell, as shown in Fig. 1. In the Fig.1. $r$ is the radius of the innermost ring and $h$ is the horizontal lattice constant, $w$ is the strip width and $d$ is the separation between the strips of the adjacent rings.

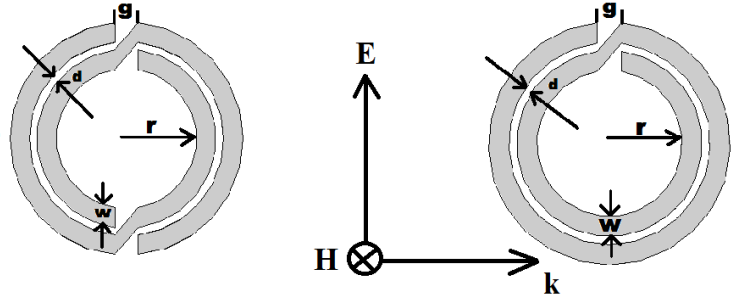

Fig.1. Schematic of Unit Cell of (a) NBSR (b) TTSR

This part is concerned with the design and experiment of SR-MNG- material (TTSR, and NBSR), so we do not elaborate the analytical model of SR variants. The numerical simulations are usually done by using TMM (Transfer Matrix Method) or commercially available simulation software. We selected the second method and designed the two structures (TTSR and NBSR) for the MNG material experiment. The numerical simulation techniques are taken from reference [10], for the unit-cell simulation where we take TTSR and NBSR (unit cell) and placed inside a parallel plate waveguide and its boundaries are PEC (Perfect Electrical Conductor) and PMC (Perfect Magnetic Conductor) respectively. From basic of theory of electromagnetic; [11] if we operate parallel plate waveguide in **TM$_0$** mode then it will be behaving like a **TEM** mode carrier or a plane wave (guide). The elaboration is as follows:

The parallel plate waveguide has the wave propagation expression as

$$\left(\frac{\partial^2}{\partial y^2} + k_c^2\right)e_z(x, y) = 0$$

where $k_c^2 = k^2 - \beta^2$ is the wave number (inside the guide). Therefore, after the solution of this equation we can get

$$e_z(x, y) = A\sin(k_c y) + B\cos(k_c y).$$

From that we can write that propagation constant $\beta = \sqrt{k^2 - k_c^2} = \sqrt{k^2 - (n\pi/d)^2}$. For $n=0$ we get $\beta = k$ (free space wave number). Thus TM$_0$ mode behaves like a TEM mode or plane wave.

The Fig.2. gives the 5-unit-cell geometry showing its top view and side view respectively.

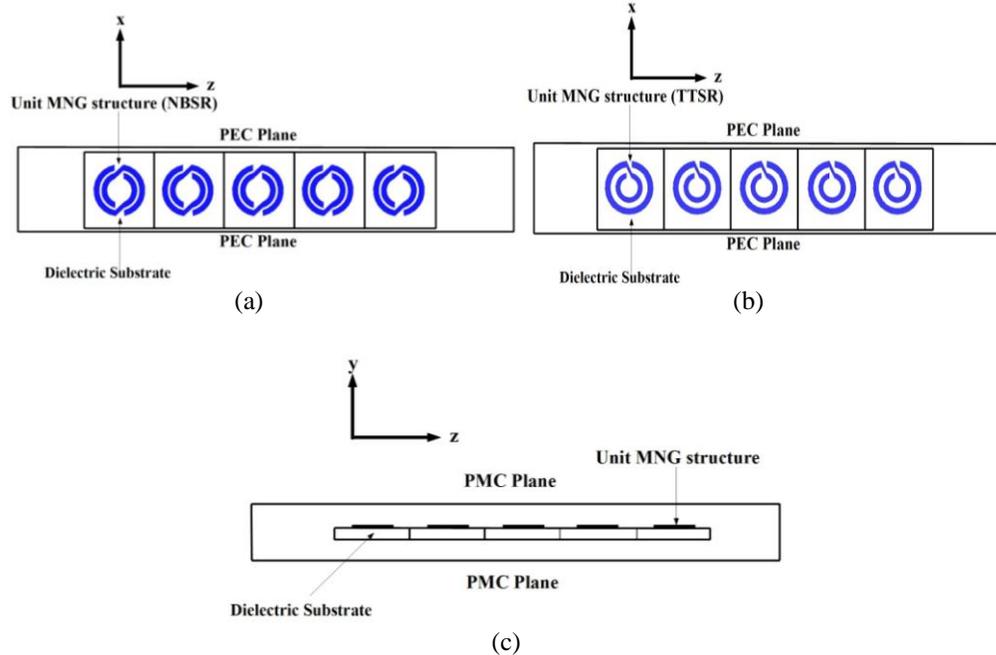

(c)

Fig.2. Planer SR-MNG materials geometry (a) NBSR and (b) TTSR Top view (c) side view of MNG-SR material.

The Table-I give the unit-cell dimensions of SR variants TTSR and NBSR. Taking this unit-cell, and results of numerical simulation we fabricated these structures and then start our experiment.

**Table: -I**

| Design parameters | TTSR | NBSR |
|---|---|---|
| No. of inclusion ($N$) | 2 | 2 |
| Inner radius/ Inner dimension ($r$) | 0.45mm | 0.85mm |
| Strip width ($w$) | 0.25mm | 0.25mm |
| Strip separation ($d$) | 0.3mm | 0.1mm |
| Gap length ($g$) | 0.3mm | 0.3mm |
| Vertical lattice constant ($v$) | 2.5mm | 2.5mm |
| Horizontal lattice constant ($h$) | 2.5mm | 2.5mm |

The experimental characterization of metamaterial is reported by many authors [12-14]. For characterization of magnetic inclusion structure the wave should be TEM (plane wave) where there is no electric and magnetic field ,present in the direction of propagation. Many researchers have done the experiment in free space [15] or some have done it in inside the waveguide [16] for characterization of metamaterials. For the sake of simplicity of experimental set up some researchers have also done the experiment in parallel plate waveguide [13,14] where the metamaterial sample is placed inside this waveguide. However the mechanism of retrieval of those experimental results, with the formulation of parameter retrieval is not detailed in the literature. The magnetic inclusion structures mounted in parallel plate waveguide having top and bottom surfaces covered with perfect electrical conductor (PEC) to eliminate the possible interferences and the side walls being open are thus emulating perfect magnetic conductor (PMC); the schematic diagram of the parallel plate experiment is given in Fig. 3.

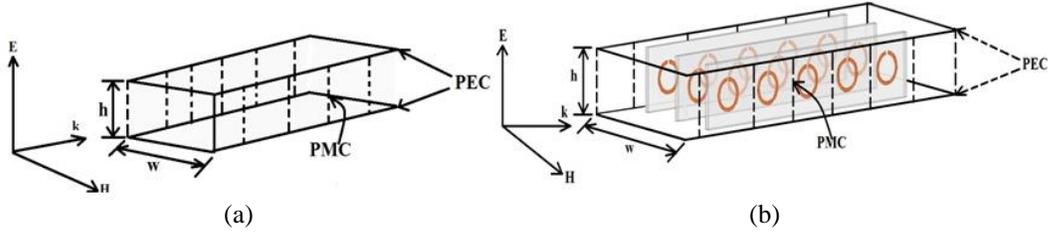

Fig.3. (a) Unloaded (b) Loaded (MNG material) for parallel plate waveguide experiments.

PEC is the material where the tangential electric field ($E_t$) is zero and its impedance is zero similarly in PMC is the material where the tangential component of the magnetic field ($H_t$) is zero and its impedance is infinity so it is called high impedance surface [17]. From the material point of view ferrite material shows the properties of PMC. For characterization of TTSR, and NBSR experimental arrangements are given in Fig. 5 respectively.

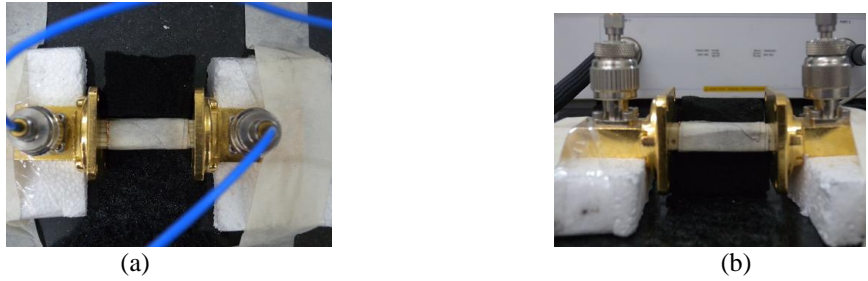

Fig.4. (a) Top view (b) Side view of the parallel plate waveguide experiment.

In the experiment the side wall is covered by the absorber of ferrite material based, prevents the outside interferences which are due to truncation of the system as we get the open boundary. To give the rigidity for the MNG sample, we have covered it with paper tape. The wave is launched by waveguide adapter (X band 8-12 GHz) which is a linearly polarized wave in *y*- directions. The direction of polarization is given in Fig.3. The two through ports are used to feed signal input in the one and the other port is terminated with the detector, for transmission experiments.

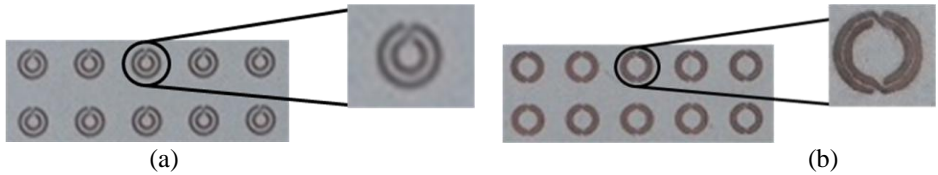

Fig.5. (a) TTSR (b) NBSR blows up view.

The TTSR, and NBSR metallic inclusion structures are imprinted on RT-Duroid 5880 (LZ) dielectric substrate in the form of a matrix (5 x 2) as shown in Fig.5. The experiment has been done by using 5 such sheets placed parallel to the narrow dimension of the parallel plate waveguide. The design dimensions (for experimental purpose) for the magnetic inclusion structures TTSR and NBSR are given in Table-I.

The results of transmission experiments with the fabricated TTSR, and NBSR structures together with the simulation results are shown in Fig. 6. (a) and (b) respectively.

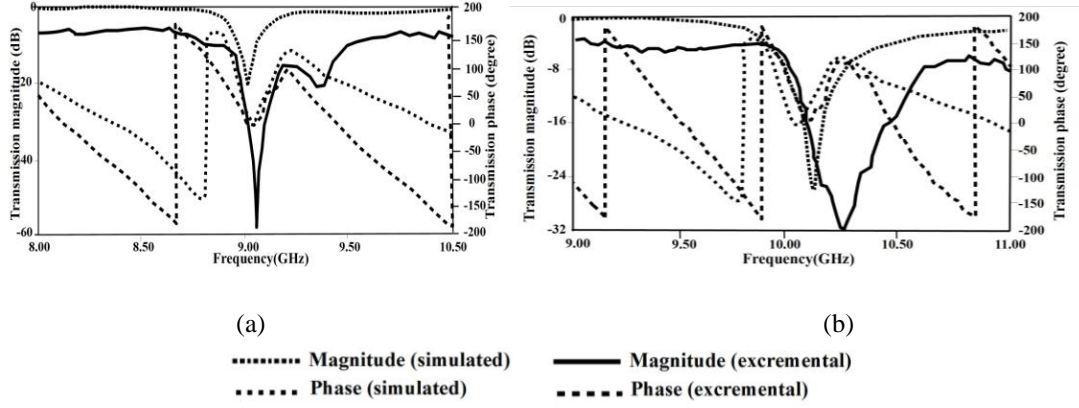

Fig.6. (a) TTSR (b) NBSR Transmission experiment and simulated results.

It is observed that the experimental result has transmission dip (stop-band) for TTSR, and NBSR is 9.10GHz, 10.25GHz, while the simulation result shows the corresponding transmission dips at 9.03GHz, and 10.10GHz.

## III. INTERPREATATIONS & DISCUSSIONS OF RESULTS

From the Fig.6. it is observed that at the MNG structure is resonating structure, in the stop band, as near the resonant frequency there is a phase change is observed. Both of NBSR and TTSR there is as a phase change of about $90^0$ is observed in both simulation and experimental data, that confirms there is some material character (permeability is negative) which goes into resonance. The general magneto static resonance is observed and the fact that MTM are resonating phenomena is confirmed. The general resonance gives opposition (to excitation) that is the direction of reaction to the action, gets reversed. Here we are observing the phase change opposite to non resonating part-confirming MTM resonated. For ENG (Epsilon Negative Material) it is the electrostatic resonance.

After getting the Scattering parameter as given in [18] the retrieved impedance and refractive index of the measured material are given in below;

$$t^{-1} = \left[\cos(nkd) - \frac{i}{2}\left(z + \frac{1}{z}\right)\sin(nkd)\right]e^{ikd} \quad (1)$$

Where $t$ = is the transmission parameter, $n$ = refractive index of the sample, $k$ = wave number ($\frac{2\pi}{\lambda}$ where $\lambda$ is wavelength (mm)), and $d$ = length of the sample (for one direction (mm)).

Similarly from the reflection coefficient [18] we can write the equation as;

$$\frac{r}{t'} = -\frac{1}{2}i\left(z - \frac{1}{z}\right)\sin(nkd) \quad (2)$$

Where the $t'$ is the normalized transmission coefficient, and $r$ = reflection coefficient, and $z$ = impedance of the sample. After some simple algebraic equation the upper two equations can be rewritten as the retrieved impedance ($z$) is;

$$z = \pm\sqrt{\frac{(1+r)^2 - t'^2}{(1-r)^2 - t'^2}} \quad (3)$$

Similarly from [18] we get a refractive index ($n$) that is;

$$\mathrm{Im}(n) = \pm \mathrm{Im}\left(\frac{\cos^{-1}\left(\frac{1}{2t'}\left[1-\left(r^2-t'^2\right)\right]\right)}{kd}\right) \qquad (4)$$

$$\mathrm{Re}(n) = \pm \mathrm{Re}\left(\frac{\cos^{-1}\left(\frac{1}{2t'}\left[1-\left(r^2-t'^2\right)\right]\right)}{kd}\right) + \frac{2\pi m}{kd} \qquad (5)$$

[Im stands for imaginary and Re stands for real part of the refractive index]

From the effective medium theory and homojunction effective medium, the external magnetic field that is perpendicular to the SR variants will cause [19];

1. Strong frequency dependent $\mu_{eff}(\omega)$.
2. Weak frequency dependent $\varepsilon_{eff}(\omega)$.

From the reference [19] the researcher has shown that the upto 11 Unicell it is an invariant of its length. From this assumption we take 5 unit cells for our parallel plate waveguide experiment effectively $d= 25$ mm (the length of the sample) there is no ambiguity on the length of the sample. Fig. 7. gives the view of the sample where the 25 mm length in the propagation direction and 10 mm in the other direction. The $d$ is smaller than the excitation wavelength (30 mm) so continuous material analysis is valid [18,19].

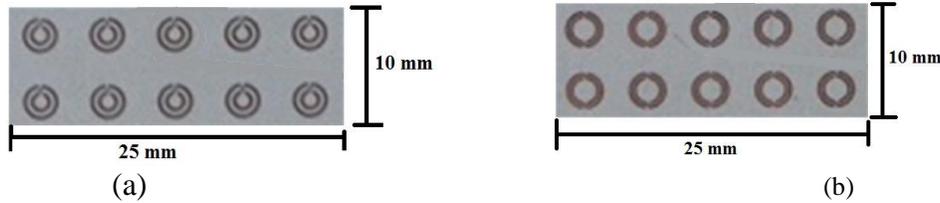

(a)  (b)
Fig.7. (a) Dimension of TTSR sample material (b) Dimension of NBSR sample material.

The Fig.8. gives the retrieve impedance ($z$) of NBSR and TTSR and Fig.9.gives the retrieved refractive index ($n$) of the metamaterial sample. For retrieved this impedance and the refractive index the values are considered $\lambda$ (wavelength) is 33 mm for TTSR and 29 mm for NBSR, the value of $k = 0.19$ mm$^{-1}$ (for TTSR) and $k = 0.21$ mm$^{-1}$ (for NBSR), similarly for $d$ the length is taken 25 mm (both for TTSR and NBSR). From the calculation of the real part of refractive index we select the term $m=3$ because for that value the $\mathrm{Re}(n_{eff}(\omega)) \approx 0$ and $\mathrm{Im}(n_{eff}(\omega)) > 0$ which corresponds that strong magnetic resonance can produce $\mu_{eff}(\omega) < 0$ [19] (and below that is imaginary for MNG). This is a criteria for working principal MTM-MNG material.

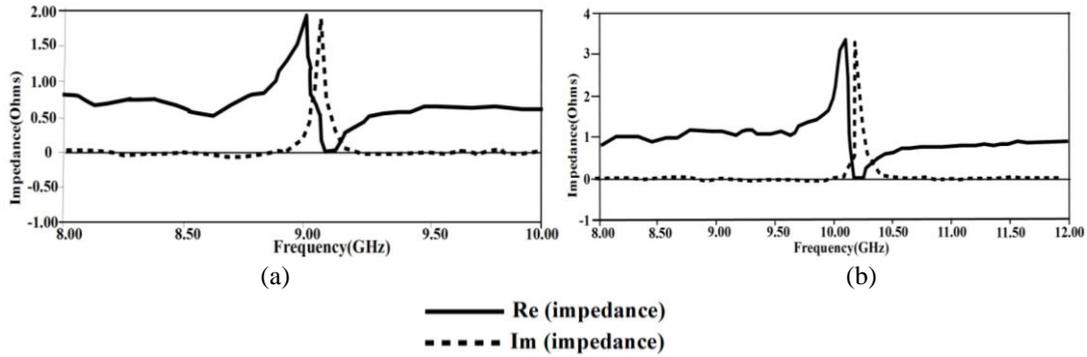

Fig.8.(a) Retrieved Impedance (z) for TTSR (real (Re) and imaginary part (Im)). (b) Retrieved Impedance (z) for NBSR (real (Re) and imaginary part (Im)).

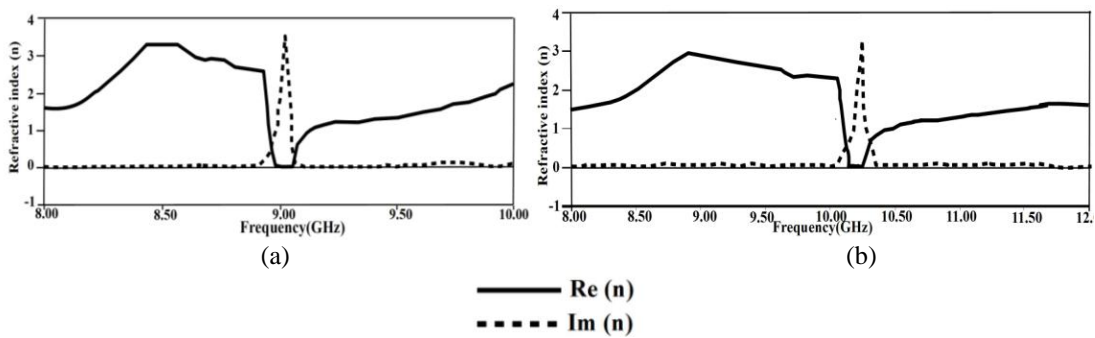

Fig.9.(a) Retrieved refractive index (n) for TTSR (real (Re) and imaginary part (Im)). (b) Retrieved refractive index (n) for NBSR (real (Re) and imaginary part (Im)).

Taking all this retrieved data (impedance ($z$) and refractive index ($n$)) the analytical, simulated, and experimental results (S-parameter retrieval method and Homojuction effective medium theory) [18,19] for TTSR, and NBSR effective relative permeability curve (simulation and experiment) are given in Fig.10 (a), and (b) respectively. The performance characteristics of TTSR, and NBSR is compared with, simulated, and experiments are given in Table: -II.

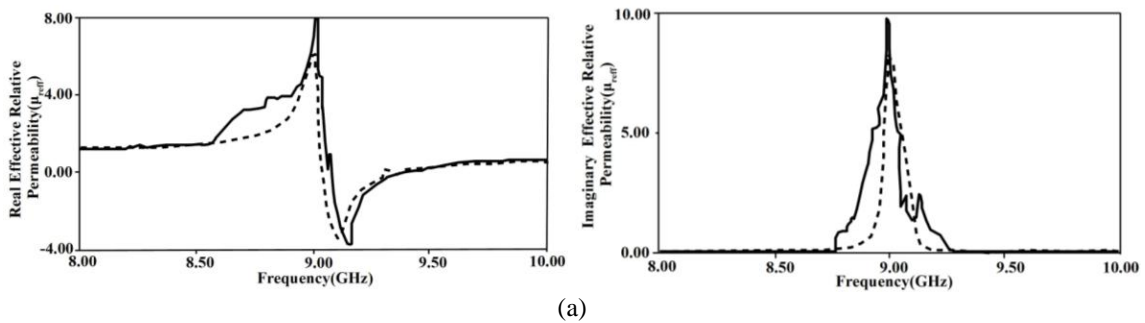

(a)

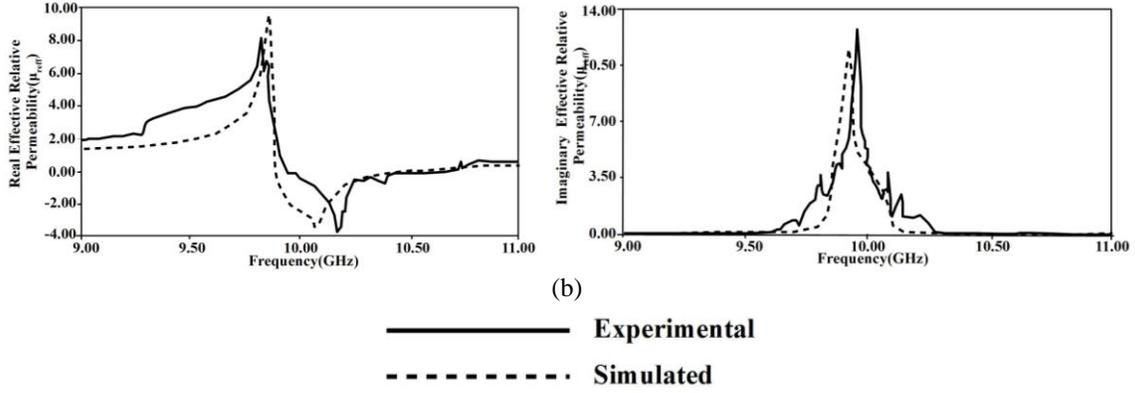

Fig.10. (a) TTSR (b) NBSR is the comparative study of numerical simulation and experimental results.

**Table:-II**

| Magnetic Inclusion Structure | $f_{m0}$(GHz) | | $f_{mp}$(GHz) | | $\Delta f = f_{mp}-f_{mp}$(GHz) | | $\mu_{reff}$ | |
|---|---|---|---|---|---|---|---|---|
| | Simu | Exp | Simu | Exp | Simu | Exp | Simu | Exp |
| **TTSR** | 9.05 | 9.08 | 9.5 | 9.55 | 0.45 | 0.43 | -3.37 | -3.7 |
| **NBSR** | 9.9 | 9.92 | 10.42 | 10.5 | 0.52 | 0.58 | -3.3 | -3.8 |

Simu:-Simulated; Exp:-Experimental results.

From the Table-II (prepared form Fig.10) we infer that in TTSR $f_{m0}$=30 MHz, $f_{mp}$ = 50 MHz and the bandwidth of negative permeability region is 20 MHz. The effective relative permeability value is same (difference in simulation and experimental results is 0.33). Similarly for NBSR we infer there is $f_{m0}$=20 MHz, $f_{mp}$ = 80 MHz and in bandwidth of negative permeability region is 60 MHz. This effective relative permeability of NBSR is similar to TTSR results (difference is 0.5).

## V. CONCULSION

Studies of different SR structures TTSR and NBSR are studied via numerical simulation and experiment with parallel plate waveguide experiments. The results are very close matched with each other, but there is slight difference is observed in experiments due to its fabrication tolerance of our present PCB design machine. In parameter retrievals we used S-parameter extraction method. Though it is a very efficient method but we found some other method to remove the ambiguity which sometimes arises in this method. This portion will discuss in our next paper.

**Acknowledgement:**

This work is fully funded by "**Board of Research in Nuclear Science"(BRNS),** Department of Atomic Energy (DAE) India, for a sanctioned project called **Left Handed Maxwell (LHM) Systems**; being carried out in SAMEER (Govt. Of India) Kolkata Centre along with collaboration with BARC. We are thankful to Dr A L Das Director SAMEER for his support and encouragement for this project.